\documentclass[12pt]{article}  

\usepackage{cite} 
\usepackage{amssymb} 
\usepackage{amsfonts} 
\usepackage{amsmath} 
\usepackage{breqn}
\usepackage{slashed} 
 
\normalbaselines
\catcode `\@=11
 
\def\section{\@startsection {section}{1}{\z@}{-3.5ex plus -1ex minus
     -.2ex}{2.3ex plus .2ex}{\normalsize\bf}}
\def\subsection{\@startsection{subsection}{2}{\z@}{-3.25ex plus -1ex minus
 -.2ex}{1.5ex plus .2ex}{\normalsize\bf}}

\def\thebibliography#1{\section*{References\markboth
  {REFERENCES}{REFERENCES}}\list
  {[\arabic{enumi}]}{\settowidth\labelwidth{[#1]}\leftmargin\labelwidth
  \advance\leftmargin\labelsep
  \usecounter{enumi}}
  \def\newblock{\hskip .11em plus .33em minus -.07em}
  \sloppy
  \sfcode`\.=1000\relax}
 
\catcode `\@=12 

\title{\bf\normalsize 
DE DONDER-WEYL HAMILTONIAN FORMULATION AND 
PRECANONICAL QUANTIZATION OF VIELBEIN GRAVITY 
 }

\author{ 
Igor V. Kanatchikov$^{}$\thanks{{\sl E-mail address}: 
{\tt kanattsi@gmail.com.}}\\
\small 
National Center of Quantum Information in Gdansk 
(KCIK),\\ 
\small 81-824 Sopot, Poland 
} 

\date{\sf\small 
} 

\newcommand{\beq}{\begin{equation}}
\newcommand{\eeq}{\end{equation}}
\newcommand{\beqa}{\begin{eqnarray}}
\newcommand{\eeqa}{\end{eqnarray}}
\newcommand{\nn}{\nonumber}

\newcommand{\pbr}[2]{ \{ \hspace*{-2.6pt} [ #1 , #2\hspace*{1.4 pt} ] 
\hspace*{-2.6pt} \} }

\newcommand{\we}{\wedge}
\newcommand{\der}{\partial}

\newcommand{\inn}{\hspace*{2pt}\raisebox{-1pt}{\rule{6pt}{.3pt}\hspace*
{0pt}\rule{.3pt}{8pt}\hspace*{3pt}}}

\newcommand{\ga}{\gamma}

\newcommand{\ka}{\varkappa}

\newcommand{\Psib}{\overline{\Psi}}
\newcommand{\Phib}{\overline{\Phi}}

\newcommand{\what}[1]{\widehat{#1}}

\newcommand{\bx}{{\mathbf{x}}}

\newcommand{\BPsi}{{\bf \Psi}} 
   
\newcommand{\BH}{{\bf H}}

\begin{document}

\maketitle

  


\begin{abstract}
\noindent
 The De Donder-Weyl (DW) covariant Hamiltonian formulation of Palatini 
first-order Lagrangian of vielbein (tetrad) gravity and its precanonical quantization are presented. No splitting into space and time is required 
in this formulation. Our recent generalization of 
 Dirac brackets is used 
to treat the  second class primary constraints appearing in the 
DW Hamiltonian formulation and to find the fundamental brackets.  
Quantization of the latter yields the representation of vielbeins  
as differential operators with respect to the spin connection coefficients 
and the 
 Dirac-like 
precanonical Schr\"odinger equation  on the space of spin 
connection coefficients and space time variables. 
The transition amplitudes on this space describe the 
quantum geometry of space-time. We also discuss the Hilbert space of the theory, 
the invariant measure on the spin connection coefficients,  
and point to the possible quantum singularity avoidance built in in
 the natural choice of the boundary conditions of the wave functions 
 on the space of spin connection coefficients. 
 
\vspace*{0.5cm}



\vspace*{0.5cm}
\end{abstract}

\section{Introduction} 

There are several dominating approaches in the literature which aim at quantization of gravity or, more generally, a synthesis of general relativity 
and quantum theory.  They can be conditionally classified according 
to their main strategies: 
\begin{itemize}
\item[1)] application of standard QFT techniques to the Lagrangians of general relativity theory or its alternatives 
(canonical QG, path integral, asymptotic safety),
\item[2)] adaptation of the classical GR to the technical requirements or limitations of QFT (LQG, shape dynamics), 
\item[3)] postulating the fundamental microscopic dynamics 
so that classical GR would appear as an effective or emergent low energy theory (string theory, GFT, induced gravity, quantum/non-commutative space-times, causal networks). 
\end{itemize} 
However, considerably less efforts have been devoted to the fourth logical possibility: 
\begin{itemize}
\item[4)]
a modification or improvement of quantum theoretic formalism and its 
adaptation to the geometric context of general relativity. 
\end{itemize} 

The distinguished role of time in the formalism and interpretation of quantum 
theory is one of the aspects to be overcome in the quantum formalism adapted 
to the goal of quantization of general relativity. As this feature of quantum description can be seen as inherited from the canonical Hamiltonian formalism 
whose structures underlie canonical quantization, one could try  
to find 
a generalization of canonical Hamiltonian formalism and its quantization in which all space-time variables would be treated on the equal footing.  
Fortunately, such generalizations are  known in the mathematical theory 
of the calculus of variations of multiple integrals and there is an 
infinite variety  of covariant finite-dimensional Hamiltonian-like formulations 
(given by different Lepage equivalents of the Poincar\'e-Cartan form \cite{kastrup,lepage,lepage1,lepage2,lepage3}) which, 
from the point of view of physics, 
implement exactly this idea.  The simplest of these formulations 
is known as the  De Donder-Weyl (DW) theory   
(see e.g. \cite{kastrup,dedonder,dedonder1}).

The DW Hamiltonian formulation of a field theory given by the first order 
Lagrangian $L = L(y^a, y^a_\mu, x^\nu) $ uses the covariant Legendre transformation to the new set of variables: {\em polymomenta} 
$$p^\mu_a := \frac{\der L}{\der y^a_\mu}$$ 
and the {\em DW  Hamiltonian function } 
$$ H(y^a, p^\mu_a, x^\mu) := y^a_\mu(y,p) p^\mu_a - L, $$ 
which, for regular theories with  
$$\det \left |\left | {\der^2 L}/ {\der y_a^\mu\der y_b^\nu }
\right |\right| \neq 0, $$
enable us to write the field equations in the 
{DW covariant Hamiltonian form:} 
\beq \label{dw}
\der_\mu y^a (x) = \frac{\der H}{\der p^\mu_a} , \quad 
\der_\mu p^\mu_a (x) = - \frac{\der H}{\der y^a} .  
\eeq
The latter  look like a multidimensional field theoretic analogue of the 
Hamilton equations with all space-time variables treated on the equal 
footing. 

A generalization of Poisson brackets to the DW Hamiltonian formulation 
\cite{mybrackets,mybrackets1,mybrackets2}, 
which is suitable for quantization, is defined on semi-basic forms on the 
polymomentum phase space (with the space-time being the base manifold and 
the space of field and polymomentum variables being the fiber) and leads 
to the Poisson-Gerstenhaber algebra structure with respect to the graded 
Lie bracket and a special $\bullet$-product of forms:   
\beq
A\bullet B:= *^{-1}(*A\we *B),  
\eeq 
called co-exterior. 

Just as the canonical quantization proceeds from the mathematical 
structures of  canonical Hamiltonian formalism, {\em precanonical 
quantization}  starts from the mathematical structures underlying 
 the  DW Hamiltonian formulation: the polysymplectic form, 
 Poisson-Gerstenhaber brackets, DW Hamilton-Jacobi theory, etc.  

It was found in our earlier work \cite{myprecanonical,myprecanonical1,myprecanonical2} 
that the quantization of the subalgebra 
of precanonically conjugate variables (similar to the Heisenberg algebra) 
leads to the following representation of the operators of polymomenta: 
\beq \label{pmuop}
\hat{p}{}^\nu_a = -i\, \hbar \ka \gamma^\nu \frac{\der}{ \der y^a},
\eeq
which act on the Clifford-valued wave functions $\Psi (y,x)$ on the 
finite dimensional covariant configuration space of field and 
space-time variables $y$ and $x$. 

The constant $\ka$ of the dimension $\ell^{\tt 1-n}$ 
in $n$ space-time dimensions appears in precanonical quantization 
on the dimensional grounds. Its meaning as the inverse 
of a very small "elementary volume" is obvious e.g. in 
the representation of the basic $(n-1)$-forms   
$$\varpi_\nu:=\der_\nu \inn (dx^0\we...\we dx^{n-1})$$ 
in terms of the space-time Clifford algebra elements:  
 $$\what{\varpi}_\nu = \frac{1}{\ka}\gamma_\nu.$$ 
Note that the approach of precanonical quantization does not modify the microscopic structure of space-time by any {\em ad hoc} assumptions. 
 
The covariant analogue of the Schr\"odinger equation in precanonical 
quantization reads 
\beq \label{nse}
i \hbar\ka \gamma^\mu\der_\mu \Psi = \hat{H}\Psi, 
\eeq
where $\hat{H}$ is the DW Hamiltonian operator composed from the partial 
differential operators with respect to the field variables, 
 c.f. (\ref{pmuop}). For the free scalar field $y$   
$$H= \frac 12 p^\mu p_\mu + \frac 12 \frac{m^2}{\hbar^2} y^2$$ 
and the operator $\hat{H}$ 
corresponds to the harmonic oscillator along the field dimension $y$ 
\cite{myprecanonical,myprecanonical1,myIJTP98a}: 
\beq \label{hscalar}
 \hat{H} = -\frac{1}{2}\ \hbar^2\ka^2 
  \frac{\der^2}{\der y^2} + 
 \frac{1}{2} \frac{m^2}{\hbar^2} y^2.  
\eeq

The self-adjointness of $\hat{H}$ with respect to the inner product 
\beq \label{scalpr}
\langle\Phi|\Psi\rangle := \int dy\ \Psib\Psi , 
\eeq 
where $\Psib:=\Psi^\dagger \ga^0$, and eq. (\ref{nse}) lead to 
the conservation law 
\beq 
\der_\mu \int dy \Psib {\gamma}^\mu \Psi =0 
\eeq
which makes the probabilistic interpretation of $\Psi(y,x)$ possible. 
Note however,  that in pseudoeuclidean space-times 
the inner product in (\ref{scalpr}) is indefinite, 
while the conserved quantity 
\beq 
\int d\bx\int dy \,\Psib\ga^0\Psi
\eeq 
is positive definite (here the notation $x^\mu = (\bx, t)$ is used). 
Hence, the approach of precanonical quantization 
implies a generalization of mathematical formalism 
of quantum theory  with an indefinite metric 
Hilbert space, where  $\gamma^0$ plays the role of  
$J$-operator (see e.g. \cite{indef}).

The particle interpretation of the free scalar field is suggested by 
the spectrum of DW Hamiltonian operator 
in (\ref{hscalar}):  
$\ka m (N+\frac{1}{2})$ with $N \in \mathbb{N}$,  
which implies  that free particles of mass $m$ correspond to the transitions 
between the neighbouring eigenstates of DW Hamiltonian operator. 

The relation   between the  
precanonical Schr\" odinger equation (\ref{nse})  
and the functional differential  Schr\" o\-dinger equation following from 
canonical quantization: 
$$i\hbar\der_t\BPsi = \what{\BH} \BPsi,$$
where $\BPsi=\BPsi$$ ([y(\bx)],t)$ is  the Schr\" odinger wave functional and $\what{\BH}$ is the functional derivative operator of the canonical Hamiltonian functional,  
 is established by assuming that there is a relation between $\BPsi$ 
 and 
 the precanonical wave function $\Psi(y,x)$ restricted to the subspace 
 $\Sigma: (y=y(\bx), t=\mathrm{const})$: 
$$\BPsi ([y(\bx)], t) = \BPsi ([\Psi_\Sigma (y(\bx), \bx, t)],[y^a(\bx)]),$$ 
 and substituting the equation for $\der_t \Psi_\Sigma$ following from 
 the restriction of (\ref{nse}) to $\Sigma$ into the 
 chain rule differentiation 
 \beq \label{dtbpsi1}
i\der_t \BPsi = 
\int d\bx~  
{\sf Tr}\left \{ 
 \frac{\delta \BPsi }{\delta\Psi^T_\Sigma(y^a(\bx),\bx, t)} 
i\der_t \Psi_\Sigma (y^a(\bx),\bx,t)  
\right \}.   
\eeq 
Then, in the limiting case ${{\gamma}^0\ka \rightarrow \delta(\mbox{\footnotesize\bf 0})}$, we are able to obtain the 
functional differential  Schr\" odinger equation as the 
consequence of  (\ref{nse})  
and the expression of the  Schr\" odinger wave functional in 
terms of the continuum product of precanonical wave functions 
\cite{my1201,my1201a} (c.f. \cite{myIJTP98,myIJTP98a}):  
\beq 
\mbox{$\BPsi$} = 
\mbox{\sf Tr} \left .\left\lbrace \prod_\bx 
 e^{-iy(\bx)\ga{}^i\der_iy(\bx)/\ka} 
 \Psi_\Sigma (y(\bx), \bx, t)
\right\rbrace{\!\!}\right\rvert_{{\gamma}^0\ka \rightarrow \delta(\mbox{\footnotesize\bf 0})}. 
\eeq 
The existence of such a relation between precanonical quantization and 
functional  Schr\" o\-dinger representation suggests that 
the standard QFT based on canonical quantization  
is a singular limit of QFT based on precanonical quantization 
when the "elementary volume" $1/\ka$ is vanishing. Note also that the 
map  ${\gamma}^0\ka \rightarrow \delta(\mbox{\bf 0})$ is actually 
the inverse of the "quantization map" 
from the exterior forms to Clifford numbers: 
$\what{\varpi}_0 = \frac{1}{\ka}\gamma_0$, which underlay precanonical quantization,  in the limit of infinite $\ka$. 

\section{DW Hamiltonian formulation of  vielbein/tetrad  gravity} 

Because the Dirac operator enters in the precanonical analogue of the covariant 
Schr\"o\-din\-ger equation (\ref{nse}), the vielbein formulation of gravity 
is a more natural  framework for precanonical  quantization than the metric formulation used in our earlier work \cite{myqg}. The latter essentially lead to a 
hybrid quantum-classical theory (c.f. \cite{elze,elze1}) because a part 
of the spin connection term in the curved space-time Dirac operator 
in (\ref{nse}) can not be expressed and quantized in terms 
of the variables of the metric formulation. 

Let us consider the  first order Palatini type 
Lagrangian density of Einstein's gravity  with the cosmological term:  
\beq \label{lagr}
{\mathfrak L}= \ \mbox{$\ \mbox{$\frac{1}{\kappa_E}$}$} {\mathfrak e} e^{[\alpha}_I e^{\beta ]}_J (\der_\alpha \omega_\beta{}^{IJ} +\omega_\alpha {}^{IK}\omega_{\beta K}{}^J) + \mbox{$\ \mbox{$\frac{1}{\kappa_E}$}$}\Lambda {\mathfrak e}, 
\eeq
where $e^\mu_I$ are the vielbein components, 
 $\omega_\alpha^{IJ}$  are torsion-free spin connection coefficients, 
$\kappa_E:= 8\pi G$, 
and ${\mathfrak e}:= \det{||e_\mu^I}||$.  

 The polymomenta associated with the vielbein  and spin connection 
 field variables treated as independent dynamical variables: 
 $${\mathfrak p}{}^\alpha_{e^I_\beta}=
\frac{\der {\mathfrak L} }{\der\ \der_\alpha e^I_\beta}  
\;\;  \mathrm{and } \;\;  
{\mathfrak p}{}^\alpha_{\omega_\beta^{IJ}} =\frac{\der {\mathfrak L} }{\der\ \der_\alpha{\omega_\beta^{IJ}}},$$ 
yield the primary constraints of the  DW Hamiltonian 
formalism, viz. 
\beq \label{constr} 
{\mathfrak p}{}^\alpha_{e^I_\beta}
\approx 0, \quad
{\mathfrak p}{}^\alpha_{\omega_\beta^{IJ}} 
\approx \ 
\mbox{$\ \mbox{$\frac{1}{\kappa_E}$}$}
{\mathfrak e} e^{[\alpha}_Ie^{\beta ]}_{J } .
\eeq 
Consequently, not all space-time gradients of vielbein and spin connection fields can  be expressed as functions of polymomenta and fields and we need to develop 
an analogue of the constraints analysis within the DW formalism. 

Notwithstanding the fact that a mathematical literature related to the 
DW Hamil\-ton\-ian theory with constraints exists \cite{dw-constr,dw-constr1,dw-constr2,dw-constr3,dw-constr4,dw-constr5,dw-constr6}, 
the  analysis suitable for  the purposes of quantization, though incomplete, 
seems  to be found only in our paper \cite{my-dirac}. The idea of that 
paper is to use the $(n-1)$-forms constructed from  the constraints 
and their Poisson-Gerstenhaber brackets found within the 
DW formalism in our earlier papers \cite{mybrackets,mybrackets1,mybrackets2},  
and to try to find a generalization of the  Dirac's treatment of constrains 
in the Hamiltonian formalism of mechanics to the DW Hamiltonian  formalism 
in field theory.  

Following this line and  using the primary constraints (\ref{constr}),  
let us write down an extended DW Hamiltonian density 
\begin{dmath}
{\mathfrak H} = 
-  \mbox{$\ \mbox{$\frac{1}{\kappa_E}$}$} 
{\mathfrak e} e^{[\alpha}_I e^{\beta ]}_J 
\omega_\alpha {}^{IK}\omega_{\beta K}{}^J  
  - \mbox{$\frac{1}{\kappa_E}$}\Lambda {\mathfrak e}
  + \mbox{$\mu_{\alpha\beta}^I {\mathfrak p}{}^\alpha_{e^I_\beta}$}
  + \lambda^{IJ}_{\alpha \beta}  \Big({\mathfrak p}{}^{\alpha}_{\omega^{IJ}_\beta}-
 \mbox{$\ \mbox{$\frac{1}{\kappa_E}$}$}
{\mathfrak e} e^{[\alpha}_Ie^{\beta ]}_{J } \Big), 
\end{dmath} 
where $\mu$ and $\lambda$  are the Lagrange multipliers. 
The DW Hamiltonian equations given by ${\mathfrak H}$ yield: 
\begin{equation} \label{dwe} 
\der_\alpha e^I_\beta = \mu^I_{\alpha \beta}, 
 \quad \der_{[\alpha} \omega_{\beta]}^{IJ} = \lambda^{IJ}_{\alpha \beta}, 
\eeq
\beq \der_\alpha {\mathfrak p}{}^\alpha_{e^I_\beta} 
= -\frac{\der{\mathfrak H}}{\der e^I_\beta}~,
\quad
 \der_\alpha  {\mathfrak p}{}^\alpha_{\omega_\beta^{IJ}} = - \frac{\der{\mathfrak H}}{\der \omega^{IJ}_\beta}  .
 \label{dwp}  
\eeq 
The first equation in (\ref{dwp}) and the second one in (\ref{dwe})
reproduce, on the constraints subspace, the Einstein equations. 
The second equation in (\ref{dwp})   and the first one in (\ref{dwe})
lead to the  covariant constancy condition:  
\beq 
\nabla{}_\beta({\mathfrak e} e^{[\alpha}_I e^{\beta ]}_{J }) =0, 
\eeq  
which can be transformed into the expression of the spin connection in terms of 
vielbeins and their derivatives. 
 
Equations (\ref{dwe}), (\ref{dwp})  are equivalent to the 
preservation of semi-basic $(n-1)-$forms constructed from 
the constraints (\ref{constr}): 
\beq
\mathfrak{C}_{e_\beta^I}:={\mathfrak p}_{e_\beta^I}^\alpha\varpi_\alpha, \quad 
\mathfrak{C}_{\omega_\beta^{IJ}}:=  
{\mathfrak p}{}^\alpha_{\omega_\beta^{IJ}} \varpi_\alpha
-
\mbox{$\ \mbox{$\frac{1}{\kappa_E}$}$}
{\mathfrak e} e^{[\alpha}_Ie^{\beta ]}_{J } \varpi_\alpha .
\eeq 
By calculating the brackets of those forms using the 
 local coordinate expression of the {\em polysymplectic form} 
 introduced in our papers \cite{mybrackets,mybrackets1,mybrackets2}:  
 \beq
 \mathbf{\Omega} = 
 d\mathfrak{p}_{e_\beta^I}^\alpha \we d {e_\beta^I} \we \varpi_\alpha 
 + d {\mathfrak p}{}^\alpha_{\omega_\beta^{IJ}} 
 \we d\omega_\beta^{IJ} \we \varpi_\alpha ,
 \eeq 
we obtain:  
 \beqa \label{cbr}
 \pbr{\mathfrak{C}_e}{\mathfrak{C}_{e'}} &=&0 , \quad \nn \\
 \pbr{\mathfrak{C}_\omega}{\mathfrak{C}_{\omega'}} &=&0 ,\quad \\
 \pbr{\mathfrak{C}_{e_\gamma^K}}{\mathfrak{C}_{\omega_\beta^{IJ}}} 
 &=& - \frac{1}{\kappa_E}\frac{\der}{\der{e_\gamma^K}}
 \left( 
 {\mathfrak e} e^{[\alpha}_Ie^{\beta ]}_{J } 
 \right)  \varpi_\alpha  , \nn 
 \eeqa
where the bracket of two semi-basic Hamiltonian 
 $(n-1)$-forms ${F}$ and ${G}$ is defined as follows: 
\beq
\pbr{{F}}{{G}}:= - X_{{F}}\!\inn d {G}, \quad 
\eeq
where 
$$X_{{F}}\!\inn \mathbf{\Omega} := d {F}. $$ 
From (\ref{cbr}) we conclude that the primary constraints 
in (\ref{constr}) are second class. 

Then, using our generalization of 
Dirac bracket 
to the singular DW Hamiltonian formalism \cite{my-dirac}:  
\beq \label{dbr}
\pbr{F}{G}{\!}^D := \pbr{F}{G} - 
 \pbr{F}{\mathfrak{C}_{U}} \bullet 
\left(\mathfrak{C}^{-1}_{UV} \we
\pbr{\mathfrak{C}_{V}}{G}\right),
\eeq
where 
the indices $U,V$ run over all components of $e$ and $\omega$ 
and the inverse of the form-valued matrix 
$\mathfrak{C}_{UV} :=  \pbr{\mathfrak{C}_U}{\mathfrak{C}_V}$ 
is defined by  
\beq \label{inverse}
\mathfrak{C}^{-1}_{UV}\we \mathfrak{C}_{VU'}:=\sigma\varpi\delta_{UU'},
\eeq  
where $\sigma=(-)1$ is the (pseudo)euclidean signature of the metric,  
 $dx^\mu\wedge\varpi_\nu =: \delta^\mu_\nu\varpi$, 
 and $\sigma\varpi\bullet$ is the unit operator when acting on any semi-basic form, 
we can obtain the following brackets of precanonically conjugate 
variables on the  subalgebra of $(n-1)$-forms: 
\beqa \label{dbr11}
{}&\pbr{{\mathfrak p}^\alpha_e \varpi_\alpha}{e' \varpi_{\alpha'}}{\!}^D=0,\\
{}&\pbr{{\mathfrak p}^\alpha_\omega \varpi_\alpha}{\omega'\varpi_{\alpha'}}{\!}^D 
 =\pbr{{\mathfrak p}^\alpha_\omega \varpi_\alpha}{\omega'\varpi_{\alpha'}}
 \,\,=\,\, \delta_\omega^{\omega'} \varpi_{\alpha'},
\label{dbr12} 
\\
{}&\!\!\hspace*{-10pt}\pbr{{\mathfrak p}^\alpha_e \varpi_\alpha}{ {\mathfrak p}_\omega}{\!}^D 
\!=\pbr{{\mathfrak p}^\alpha_e \varpi_\alpha}{\omega' \varpi_{\alpha'}}{\!}^D 
\!=\pbr{{\mathfrak p}^\alpha_\omega\varpi_\alpha}{e' \varpi_{\alpha'}}{\!}^D \!=0,\label{dbr13}
\eeqa 
and similar precanonical brackets of $(n-1)$- and $0$-forms, which also 
 constitute a subalgebra with respect to the Poisson-Gerstenhaber bracket operation: 
\beqa \label{dbr11b}
{}&\pbr{{\mathfrak p}^\alpha_e \varpi_\alpha}{e'}{\!}^D=0,\\
{}&\pbr{{\mathfrak p}^\alpha_\omega \varpi_\alpha}{\omega'}{\!}^D 
 =\pbr{{\mathfrak p}^\alpha_\omega \varpi_\alpha}{\omega'} \,\,=\,\, \delta_\omega^{\omega'},
\label{dbr12b} 
\\
{}&\!\!\hspace*{-10pt}\pbr{{\mathfrak p}^\alpha_e \varpi_\alpha}{ {\mathfrak p}_\omega}{\!}^D 
\!=\pbr{{\mathfrak p}^\alpha_e \varpi_\alpha}{\omega}{\!}^D 
\!=\pbr{{\mathfrak p}^\alpha_\omega\varpi_\alpha}{e}{\!}^D \!=0, \label{dbr13b}
\eeqa 
and 
\beqa \label{dbr11c}
{}&\pbr{{\mathfrak p}^\alpha_e}{e' \varpi_{\alpha'}}{\!}^D=0,\\
{}&\pbr{p^\alpha_\omega}{\omega'\varpi_\beta}{\!}^D=
\pbr{p^\alpha_\omega}{\omega'\varpi_\beta} = \delta^\alpha_\beta \delta^\omega_{\omega'},
\label{dbr12c} 
\\
{}&\!\!\hspace*{-10pt}\pbr{{\mathfrak p}^\alpha_e }{ {\mathfrak p}_\omega \varpi_{\alpha'}}{\!}^D 
\!=\pbr{{\mathfrak p}^\alpha_e }{\omega \varpi_{\alpha'}}{\!}^D 
\!=\pbr{{\mathfrak p}^\alpha_\omega}{e' \varpi_{\alpha'}}{\!}^D \!=0.\,\label{dbr13c}
\eeqa 
 
The following remarks regarding the above calclulation are in order. 
Note that the formula in (\ref{dbr}) assumes that 
$\mathfrak{C}^{-1}_{UV}$ exists in the sense of (\ref{inverse}) 
\cite{my-dirac}. 
However, it is not the case for the matrix 
defined by (\ref{cbr}): 
\beq \label{cuv}
\mathfrak{C}_{UV} := 
\begin{Vmatrix}
0 & \mathfrak{C}_{e\omega} \\
\mathfrak{C}_{\omega e} & 
0
\end{Vmatrix},
\eeq
where $\mathfrak{C}_{e\omega} := \pbr{\mathfrak{C}_{e}}{\mathfrak{C}_{\omega}}$  
is a rectangular matrix (16$\times$24 in $n=4$ dimensions). In the usual Dirac's Hamiltonian formalism it would signal that not all of the second class constraints are found. However, it is not necessarily the case here, because the number of polymomenta is different 
from the number of field variables and the analogues of the symplectic matrix 
in the polysymplectic formalism  are singular matrices similar to the higher dimensional Duffin-Kemmer-Petiau matrices (c.f. \cite{dkp}) whose algebraic definition is actually tantamount to the statement that, up to a sign factor, 
they are generalized Moore-Penrose inverse to themselves. 
More generally than in  (\ref{inverse}) we can understand 
$\mathfrak{C}^{-1}_{UV}$ as a generalized inverse such that 
\beq 
 \mathfrak{C}_{UU'} \bullet
(\mathfrak{C}^{-1}_{U'V'}\we \mathfrak{C}_{V'V})
 =  \mathfrak{C}_{UV} .
\eeq 
Then the specific structure of (\ref{cuv}) ensures that the  Moore-Penrose-type generalized inverse of $\mathfrak{C}_{UV}$ has the same 
matrix block structure  
   as (\ref{cuv}) with $\mathfrak{C}_{e\omega}$ replaced 
   by $\mathfrak{C}^{-1}_{e\omega}$,  
   so that 
(\ref{inverse}) is fulfilled 
 on the $e$-subspace, viz. 
\beq \label{e-inverse}
\mathfrak{C}^{-1}_{eU'}\we \mathfrak{C}_{U'e'} 
= \delta_{ee'} \sigma \varpi .
\eeq 
This is the $e$-subspace inverse which is needed in 
order to calculate the 
brackets in (\ref{dbr11}), (\ref{dbr11b}), and (\ref{dbr11c}). 
For example, 
by denoting ${\mathfrak{p}_e}:={\mathfrak{p}_e^\alpha\varpi_\alpha}$, 
we obtain: 
\begin{equation}
\pbr{e'}{\mathfrak{p}_e}{\!}^D = 
\pbr{e'}{\mathfrak{p}_e}
- \pbr{e'}{\mathfrak{C}_{e''}}
\bullet\left(\mathfrak{C}^{-1}_{e''\omega} \we\pbr{\mathfrak{C}_\omega}{\mathfrak{p}_e}\right) = -\delta_{e'}^e + \delta_{e'}^{e''}\bullet 
\left(\mathfrak{C}^{-1}_{e''\omega}\we \mathfrak{C}_{\omega e}\right) = 0, 
\end{equation} 
which is the result in (\ref{dbr11b}).
All other brackets 
in (\ref{dbr11})-(\ref{dbr13c})
are vanishing as a consequence of the specific matrix block structure of 
$\mathfrak{C}_{UV}$  and its generalized inverse. 

The brackets in (\ref{dbr11})-(\ref{dbr13c}) are assumed to be 
the analogue of the fundamental Dirac brackets of canonical variables in constrained mechanics and they underlie the quantization procedure below.

\section{Quantization} 

Usually, quantization of systems with second class constraints is performed by transforming the Dirac brackets into commutators according 
to the Dirac's quantization rule. However, in the present approach 
the latter has to be modified in order to make sure that densities 
are quantized as density valued operators, viz. 
\beq \label{dirule}
[\hat{A}, \hat{B}]= 
- i\hbar \what{\mathfrak{e}\pbr{A}{\!B\!}{\!}^D}, 
\eeq
where (the operator of) $\mathfrak{e}$ appears due to the fact that the polysymplectic form and polymonenta are densities.

Now,  quantization of brackets in (\ref{dbr11}), (\ref{dbr13}), 
(\ref{dbr11b}), (\ref{dbr13b}), (\ref{dbr11c}), (\ref{dbr13c})    
and the constraint ${p}_e\approx 0$ lead us to the conclusion that 
 the operators of the conjugate polymomenta of vielbeins are zero: 
 $\hat{p}_e = 0$. 
We can, therefore, set our precanonical wave function to  depend 
only on the spin connection and space-time variables, i.e.  $\Psi=\Psi(\omega^{IJ}_\alpha, x^\mu)$.

Further, quantization of the Dirac bracket in (\ref{dbr12b}) yields 
\beq
\what{p^\alpha_{\omega^{IJ}_\beta}\varpi_\alpha} = 
- i\hbar\hat{\mathfrak{e}}\,\frac{\der}{\der \omega^{IJ}_\beta}.
\eeq
Moreover, by quantizing (\ref{dbr12c}), 
which coincides with the familiar bracket of polymomenta and field variables  
that  underlies precanonical quantization in flat space-time 
\cite{myprecanonical,myprecanonical1},  
we obtain the formal representation of polymomenta: 
\beq \label{ptheta}
\hat{{\mathfrak p}}{}^{\alpha}_{\omega_\beta^{IJ}} 
 = -i \hbar\ka\,\mbox{\raisebox{-2pt}{$\vdots$}}
\hat{\mathfrak e}\,\hat{\gamma}{}^{[ \alpha} 
\frac{\der}{\der \omega_{\beta]}^{IJ}} 
\mbox{\raisebox{-2pt}{$\vdots$}}~,  
\eeq 
where the density $\hat{\mathfrak e}$ and 
the 
curved space-time Dirac matrices $\hat{\gamma}{}^{\alpha}$ are yet unknown operators,  and  \mbox{\raisebox{-2pt}{$\vdots$}} 
stands for a potential operator ordering ambiguity. 
Note that when obtaining  (\ref{ptheta})  we still assumed that 
\beq \label{omnuop}
\what{\varpi}_\nu = \frac{1}{\varkappa}\what{\gamma}_\nu,
\eeq
which is just  a formal generalization of the relation known from 
precanonical quantization in flat space-time \cite{myprecanonical,myprecanonical1},  
as far as the explicit operator representation of $\what{\gamma}_\nu$ is not 
known.

Next, let us insert the precanonical operator representation of 
$\hat{{\mathfrak p}}{}^{\alpha}_\omega$, 
eq. (\ref{ptheta}), 
into the  strong operator version of the second constraint in (\ref{constr})  
and contract it with  flat  $\bar{\gamma}^{IJ}$-s:
\beq \label{c1}
{({\mathfrak e} e^{[\alpha}_I e^{\beta ]}_J \bar{\gamma}^{IJ}}){}^{op} = 
\what{{\mathfrak e} \gamma}^{\alpha\beta} =  {\kappa_E ({\mathfrak p}  {}^\alpha_{\omega_\beta^{IJ}}\bar{\gamma}^{IJ})
}{}^{op}, 
\eeq 
where $()^{op}$ replaces the hat over the longer expressions 
and 
$$\hat{\gamma}{}^\nu := \hat{e}{}^\nu_I \bar{\gamma}^I,$$ 
where $\bar{\gamma}^I \bar{\gamma}^J + \bar{\gamma}^J\bar{\gamma}^I = 
2 \eta^{IJ}$, $\eta^{IJ}$ is 
a fiducial flat Minkowski  metric with the signature $+--...-$, 
and $\bar{\gamma}^{IJ} := \bar{\gamma}^{[I}\bar{\gamma}^{J]}.$
A comparison with (\ref{ptheta}) yields the operator representation of 
the 
curved space-time Dirac matrices:  
\beq
 \what{\gamma}{}^\beta = -i \hbar\ka\kappa_E \bar{\gamma}^{IJ}\frac{\der}{\der \omega_{\beta}^{IJ}}~,  \label{gamma} 
\eeq
vielbeins: 
\beq 
\hat{e}{}^\beta_I = -i \hbar\ka\kappa_E \bar{\gamma}^{J}\frac{\der}{\der \omega_{\beta}^{IJ}}~, \label{eop}
\eeq
and 
the 
polymomenta conjugate to  spin connection:
\beq \label{opom}
\hat{{\mathfrak p}}{}^{\alpha}_{\omega_\beta^{IJ}} 
 = - \hbar^2\ka^2\kappa_E 
\,\hat{\mathfrak e}\,
\bar{\gamma}^{KL}\frac{\der}{\der \omega_{[\alpha}^{KL}}
\frac{\der}{\der \omega_{\beta]}^{IJ}} 
~,  
\eeq
where the operator of $\mathfrak{e}$ can now be constructed from 
(\ref{eop}):
 \beq \label{eeop}
\hat{\mathfrak{e}}{} = 
\left( \frac{1}{n!} \epsilon^{I_1...I_n}\epsilon_{\mu_1...\mu_n} 
\hat{e}{}^{\mu_1}_{I_1} ... \hat{e}{}^{\mu_n}_{I_n} \right)^{-1} . 
\eeq 
We can also obtain the operators of $(n-1)$-volume elements (\ref{omnuop}),
that leads to a rather complicated non-local expression:  
\beq
\what{\omega}_\nu = \frac{1}{\ka (n-1)!} \ \hat{\mathfrak{e}} \
 \epsilon_{\nu\mu_1 ... \mu_{n-1}}
\what{\gamma}{}^{\mu_1} ... \what{\gamma}{}^{\mu_{n-1}} , 
\eeq 
and the operator of the metric tensor $g^{\mu\nu}$:  
\beq
\what{g^{\mu\nu}} = -\hbar^2 \ka^2\kappa_E^2 
\eta^{IJ}\eta^{KL}\frac{\der^2}{\der\omega^{IK}_\mu \der\omega^{JL}_\nu}~.
\eeq 

Finally, using (\ref{opom}) we construct the DW Hamiltonian 
operator $\what{H}$ which corresponds to the DW Hamiltonian density 
restricted to the subspace of constraints (\ref{constr}), 
${{\mathfrak e} H}:=  {{\mathfrak H}}|_C$: 
\begin{equation} \label{hgrop}
\what{H} = \hbar{}^2\ka^2\kappa_E \bar{\gamma}^{IJ} 
\mbox{\raisebox{-2pt}{$\vdots$}}\, 
\omega_{[\alpha}{}^{KM}\omega_{\beta] M}{}^L 
\frac{\der}{\der \omega_{\alpha}^{IJ}} \frac{\der}{\der \omega_{\beta}^{KL}} 
\mbox{\raisebox{-2pt}{$\vdots$}}  
- \frac{1}{\kappa_E} \Lambda~.
\end{equation}

\section{Covariant  Schr\"odinger equation for quantum gravity}

The precanonical covariant  Schr\"odinger equation which generalizes eq. 
(\ref{nse})  to the context of quantum gravity will have the form 
\beq \label{nsepsi}
i \hbar\ka 
\what{\slashed\nabla}  \Psi = 
\what{H} \hspace*{-0.0em} \Psi,   
\eeq
where
$
\what{\slashed\nabla} 
:= 
({{\gamma}^\mu(\der_\mu+\omega_\mu)})^{op}$  
  with the spin connection term  
  $\omega_\mu := \frac{1}{4} \omega_{\mu IJ} \bar{\gamma}^{IJ}$ 
~ is what we called the "quantized Dirac operator", because 
the Dirac matrices and the spin connection term in it are now operators 
themselves. Using the operator representation 
of the curved space ${\gamma}$-matrices in (\ref{gamma}) we obtain: 
\beq 
\what{\slashed\nabla}  = 
-i \hbar\ka\kappa_E \bar{\gamma}^{IJ} 
\mbox{\raisebox{-2pt}{$\vdots$}} 
\frac{\der}{\der \omega_{\mu}^{IJ}} 
\left(\der_\mu +  \frac{1}{4} \omega_{\mu KL}\bar{\gamma}^{KL}\right) 
\mbox{\raisebox{-2pt}{$\vdots$}}~.
\eeq
Therefore, the precanonical counterpart of the Schr\"odinger 
equation for quantum gravity takes  the form 
\begin{dmath} \label{wdw}
\bar{\gamma}{}^{IJ} 
\mbox{\raisebox{-2pt}{$\vdots$}} 
 \left( \der_\mu +   \frac{1}{4} \omega_{\mu KL}\bar{\gamma}^{KL} 
  - 
  \omega_{\mu M}^{K}\omega_{\beta}^{ML} 
  \frac{\der}{\der \omega_{\beta}^{KL}} 
  \right) 
\frac{\der}{\der \omega_{\mu}^{IJ}}   
  \mbox{\raisebox{-2pt}{$\vdots$}}
   \Psi 
       + \,  \frac{\Lambda}{\hbar^2\ka^2\kappa_E^2} \Psi  = 0  
\end{dmath} 
and determines  the   wave function $\Psi(\omega,x)$ or, 
more generally,  the  transition   amplitudes
$\langle \omega,x |\omega',x' \rangle$. 
The latter  provide an inherently  quantum description of the 
geometry of space-time which generalizes  classical 
differential geometry with its smooth connection fields $\omega(x)$. 

Let us note that the combination of the constants including $\ka$ 
in the last term of (\ref{wdw}) is dimensionless. By fixing the 
 operator ordering in (\ref{wdw}) we would generate a dimensionless 
 constant of the order $\sim n^6$ (the number of components of $\omega$-s 
 is $\sim n^3$) which can be interpreted as the cosmological constant 
 $\Lambda$ devided by ${\hbar^2\ka^2\kappa_E^2}$. In this case, however, 
 the observable value of $\Lambda$ is obtained (at $n=4$) only if 
 $\ka$ is roughly at the 
 nuclear energy scale, which is far away from our original expectation  
 that $\ka$ is at about the Planck scale and contradicts the experimental evidence that the usual relativistic space-time holds even at  {\em TeV} scale. 
On the other hand,  if we take $\ka$ at the Planck scale then we arrive 
at the familiar 120 orders of magnitude error in the estimation of the cosmological constant, which is usually obtained by using the Planck scale 
cutoff in the momentum space integration of the zero point energies. 
This rather confirms that the constant $\ka$ of precanonical quantization is related to the ultra-violet cutoff scale and that the cosmological constant problem is not related to the ground state of pure quantum gravity but 
rather to the particle composition of the universe.  


\section{Hilbert space}

It is natural to assume that the wave functions $\Psi(\omega, x)$
vanish at large values of $\omega$-s.    
Then the probability amplitude of observing the regions of space-time 
with very large curvature 
 is very small,  so that the quantum gravitational singularity avoidance 
is essentially built in in the choice of the boundary condition in 
 $\omega$-space. 

The scalar product is expected to have the form: 
$$
\left\langle \Phi | \Psi \right\rangle 
:=  \int [d\omega] \Phib\Psi, 
$$
where $[d\omega]$ is an invariant measure on the space of 
spin connection coefficients. 
Using the arguments similar to those used by Misner to obtain the invariant measure on the space of metrics \cite{misner}, we found: 
\beq 
[d\omega]={\mathfrak e}{}^{- n(n-1)}\prod_{\mu, I<J} d \omega_\mu^{IJ}. 
\eeq
Because in the present picture ${\mathfrak{e}}$ is an operator given by 
(\ref{eeop}),  the measure  $[d\omega]$ is operator valued 
and the scalar product of the theory has the form 
\beq 
\left\langle \Phi | \Psi \right\rangle 
:=  \int \Phib \, \what{[d\omega]}_{} \Psi.
\eeq
Then  the most natural definition of the expectation values of operators 
using the scalar product with the operator valued measure implies 
the Weyl ordering, viz. 
\beq \label{weo}
\langle \what{O}\rangle  := \int \Psib \, \left(\what{[d\omega]}\what{O}\right)_{W} \Psi. 
\eeq

When discussing the specific physical problems using the formalism of this 
paper we will have to distinguish between the physical aspects and those 
attributed to the choice of coordinates. The latter are a macroscopic 
notion due to the observer's choice and, therefore, can be implemented 
on the average. For example, the choice of the harmonic coordinates on 
the average  leads to the following condition on the wave  function $\Psi(\omega,x)$:  
\beq \label{ddf} 
\der_\mu \left\langle \Psi(\omega,x) \left| \what{\mathfrak{e} g^{\mu\nu}}
\right|\Psi(\omega,x) \right\rangle = 0  ,
\eeq  
which should be solved together with the covariant Schr\"odinger equation, eq. (\ref{wdw}), 
and that makes the problem more complicated.

\section{Conclusion}

Quantization of vielbein gravity using the approach of precanonical 
quantization, which is based on the  De Donder-Weyl  covariant Hamiltonian formulation, is discussed. All space-time variables are treated on the 
equal footing as generalizations of the time parameter in non-relativistic mechanics.  No global splitting  to space and time is required.  

The DW  Hamiltonian formulation of the first order Palatini action of 
vielbein gravity leads to the second class primary constraints which  
are treated using  our recent generalization of 
Dirac bracket 
to the DW formalism \cite{my-dirac}. The consideration of the fundamental 
generalized Dirac brackets of precanonically conjugate variables, 
eqs. (\ref{dbr11})-(\ref{dbr13c}), leads  to the conclusion that the quantum dynamics of gravity can be formulated using the wave functions on the space of spin connection coefficients and space-time variables. 
The operators of vielbeins,  metric tensor, 
DW Hamiltonian operator and the quantized Dirac operator which enters  
the covariant precanonical Schr\" odinger equation of quantum gravity 
are explicitly constructed.  We also discuss the Hilbert space of the theory 
and the invariant operator-valued integration measure on the space of spin connection coefficients. 
Let us note  that the resulting (still tentative) formulation 
of quantum gravity is non-perturbative, covariant and background independent.


However, it is  not clear at this stage if the consideration of the fundamental Dirac brackets in (\ref{dbr11})-(\ref{dbr13c}) is sufficient, because on 
the subalgebra of $0$- and $(n-1)$-forms we can also calculate brackets between the forms composed from vielbeins and spin connection coefficients,  such as  
$\pbr{e}{\omega\,\varpi_\mu}{\!}^D  \sim \der_\mu \inn \mathfrak{C}^{-1}_{ew},$ 
which explicitly depend on  the complicated nonlinear expression of the 
generalized inverse of the rectangular matrix $\mathfrak{C}_{ew}$ 
in  (\ref{cbr}) and cannot be quantized directly using the 
Dirac's quantization rule (\ref{dirule}). 

Among the issues left beyond the scope of the paper there are 
the details of the 
choice of the coordinate (gauge) conditions on the average (c.f. \cite{previous}), 
which are not sufficiently clear to us,  
and the issues related to the indefinite inner product Hilbert space 
appearing in the formalism of the theory. We were also unable so far to demonstrate that the present formulation reproduces the Einstein equations 
on the average or in the classical limit.  We hope to elaborate on those 
issues in the forthcoming publications.

\subsection*{Acknowledgments} 
{I thank R. Sverdlov for his critical remarks 
and M. Reginatto and A. Khugaev for encouraging discussions. }


\end{document}